\documentclass[12pt,a4paper]{article}

\usepackage{times} 
\usepackage{cite}

\title{\huge \bf Dimesoatoms production in high energy collisions}
\author{L. Afanasyev, S. Gevorkyan, O. Voskresenskaya}
\date{}

\begin{document}
\maketitle

\begin{center}
Joint Institute for Nuclear Research, Dubna, Moscow Region, 141980 Russia
\end{center}

\begin{abstract}
The production of two meson electromagnetic bound states and free meson pairs
$\pi^+\pi^-$, $K^+K^-$, $\pi^+ K^{\mp}$ in relativistic collisions
has been considered. It is shown that making use of the exact Coulomb wave function for
dimesoatom (DMA) allows one to calculate the yield of any $n$S state with desired accuracy.
 The relative probabilities of production of DMA and meson pairs in the free state are estimated.
 The amplitude of DMA transition from 1S to 2P state, which is essential for the pionium Lamb
 shift measurements, has been obtained.

\end{abstract}

\section{Introduction}

More than fifty years ago it was shown that the low energy scattering
properties of strongly interacting charged particles are  related to
the hadronic properties of hydrogenlike atoms formed by such particles \cite{uretsky61}. In
particular, the investigation of the ground state lifetime $\tau_0$ of hadronic atoms and
their Lamb shifts allows  one to determine scattering lengthes of meson-meson scattering~\cite{nemenov85}.
There were considered all details of experiment on observation and study of such atoms produced in inclusive high-energy interactions.
An approach for DMA ground state lifetime determination based on description of DMA as a
multilevel system propagating and interacting in a target was developed in \cite{afan96}.
For the first time, experiment on the $\pi^+\pi^-$ atoms study was done at Protvino at U-70
accelerator \cite{afan93} and then continued at Proton Synchrotron at CERN ~\cite{dirac95},
where the lifetime of the DMA ground state $\tau_0$ has been obtained~\cite{adeva11}.
This lifetime is dominated by the annihilation process $\pi^+\pi^-\to\pi^0\pi^0$ and predicted very accurately in the Chiral Perturbation Theory~\cite{Wein79,gasser83,Gasser85,Colan01NP}. Recently,
new data on production of $\pi^+ K^{\mp}$ atoms~\cite{adeva14}
and first observation of long-lived pionium atoms have been reported~\cite{adeva15}.

The facts mentioned above have  stimulated us to consider in detail the
production of bound and unbound states for any mesons.
We start (in Sec.~2) from a consideration of the bound and unbound meson pairs creation in
inclusive processes and obtain an expression, which connects the inclusive production cross sections of bound and unbound states of the interacting meson pair with the double inclusive production cross section of noninteracting pair at zero relative momentum.
We also compare  the probability  of meson pairs production in discrete and continuous  states
(Sec.~3) and estimate the accuracy of ``zero radius'' approximation, which widely exploited
in such type considerations (Sec.~4). Then, in Sec.~5, we derive an amplitude  and a relevant differential cross section for  DMA transition from 1S to 2P state in an analytic form.
This transition is the main source of the 2P states  that is crucial for the Lamb shift measurements~\cite{nemenov01,nemenov02} in experiments with dimesoatoms. Finally, in Sec.~6, we summarize our results. The Appendix contains some details of the evaluation of integral (\ref{eq:int}).

\section{Dimesoatoms creation in inclusive processes}
Let us consider a process of producing a pair of two  oppositely charged mesons
$h^+,h^-$ in the inclusive process

\begin{equation}
 a+b\to h^+ + h^- + X\,,
\label{eq:react}
\end{equation}
where $X$ are  particles whose momenta distribution is not essential in the later on consideration.
in the further consideration
The matrix element of the process (\ref{eq:react})
\begin{equation}
M\left(\vec p_+,\vec p_-\,;\{\vec p_x\}\right)
\end{equation}
is a function of $h^+, h^-$, momenta $p_+,p_-$,  and the momenta of accompanying particles {$p_x$}.
The invariant distribution of hadrons $h^+, h^-$ has a usual form

\begin{eqnarray}
&&(2\pi)^6\,2E_+\cdot 2E_-\frac{d\sigma}{d\vec p_+d\vec p_-} =
\frac{1}{4\sqrt{(p_ap_b)^2-m_a^2m_b^2}}\nonumber\\
&&\times  \int \left| M(\vec p_+,\vec p_-\,;\{\vec p_x\})\right|^2
(2\pi)^4\delta^{4}(p_a+p_b-p_+-p_--{p_x})d\Phi_x\,, \nonumber\\
\label{eq:cs}
\end{eqnarray}
where the phase space volume reads

\begin{equation}
 d\Phi_x=\prod^{N(X)}_{i=1} \frac{d^3\,\vec p_i}{(2\pi)^3\,2E(\vec p_i)}\,.
\end{equation}

Introducing  the total and relative momenta of hadrons $h^+, h^-$

\begin{eqnarray}
\vec P&=&\vec p_++ \vec p_-\,,\nonumber\\
\vec p&=&\frac{\mu}{m_+}\vec p_+ -\frac{\mu}{m_-}\vec p_-\,,
\end{eqnarray}
where $\mu=m_+ m_-/(m_+ +m_-)$ is the reduced mass of $h^+, h^-$,
 the invariant matrix element can be rewritten through new independent variables:

\begin{equation}
 M\left(\vec p_+,\vec p_-\,;\{\vec p_x\}\right)\rightarrow M(\vec P,\vec p\,;\{\vec p_x\})\,.
\end{equation}

In the rest frame of $h^+h^-$ pair the total momentum {$\vec{P}=0 $} and the amplitude
 $M(\vec{P},\vec{p};\{\vec{p}_x\})$ would be a monotone function of relative
momentum $p$ if the interaction in the $h^+h^-$ pair is absent. The
interaction in the final state (in particular, the coulomb attraction
between $h^+$ and $h^-$) violate the monotone behavior of the matrix
element at small relative momenta $p\leq \alpha \mu$
($\alpha=e^2/4\pi=1/137$ is the fine structure constant) and
leads to creation of coupled states of $h^+h^-$ system ($h^+h^-$
atoms).

To obtain the production amplitude of the $h^+h^-$ system in
any certain state $f$, one has to project the amplitude of the
pair of noninteracting hadrons $M_0(\vec P\,,\vec p\,;$ $\{\vec p_x\})$ on
this state

\begin{equation}
R_f=\int M_0\left(\vec 0,\vec p\,;\{\vec p_x\}\right)\psi_f(\vec p)d^{\,3} p\,,
\end{equation}
where  $\psi_f(\vec p)$ is a wave function of the state  $|f\rangle$  in the momentum representation.
 The  wave function in the coordinate representation can be obtained by Fourier transformation
\begin{equation}
 \psi_f(\vec p)= \frac{1}{(2\pi)^{3/2}}\int \psi_f(\vec r)e^{i\vec p\,\vec r}d^{\,3}r
\end{equation}
 with the normalization
\begin{equation}
 \int \psi_{f^{\prime}}(\vec r)\psi_{f}(\vec r)d^{\,3} r=\delta_{ff^{\prime}}\,.
\end{equation}
The Kronecker symbol  $\delta_{ff^{\prime}}$ would  be interpreted as a product
$\delta_{ff^{\prime}}=\delta_{nn^{\prime}}\delta_{l\,l^{\,\prime}}\delta_{mm^{\prime}}$ when the states
 $|f\rangle$, $|f^{\prime}\rangle$ belong to the discrete spectra with a set of quantum numbers
$n, l, m $ (main, orbital, and magnetic quantum  numbers, respectively) or as a product
$\delta_{ff^{\prime}}=\delta(k-k^{\prime})\delta_{l\,l^{\,\prime}}\delta_{mm^{\prime}}$
when the  states  $|f\rangle$, $|f^{\prime}\rangle$ belong to continuous spectra with the wave number $k$ instead of the main quantum number.

The analysis of  the  $R_f$ dependence on quantum numbers of state $|f\rangle$  is more transparent in a coordinate representation.
Introducing the  relevant  amplitude for the production of noninteracting hadrons in the final state
\begin{equation}
M_0(\vec r)=\frac{1}{(2\pi)^{3/2}}\int M_0(\vec p)  e^{i\vec p\:\vec r}d^{\,3} p\,,
\end{equation}
where $M_0(\vec p)\equiv M_0(\vec 0,\vec p\,;\{\vec p_x\})$, one gets
\begin{equation}
R_f=\int M_0(\vec r)\psi_f(\vec r)d^{\,3} r\,.
\label{eq:rf}
\end{equation}
Since the production of $h^+h^-$ pairs is a result of strong interaction
which decreases exponentially, the amplitude $M_0(\vec r)$ is nonzero
only at small distances $r\leq 1/m $. As
for the distances, where the wave
function $\psi_f(\vec r)$ changes essentially, they are of the order of Bohr
radius $r_B=1/\mu\alpha$ for bound state and of the order of $1/k$ for
continuous states in the case of pure electromagnetic interactions.
Thus, we can take out the slowly varying wave function at $\vec r=0$
and put it in front of the integral (\ref{eq:rf}). So we obtain
\begin{equation}
R_f=\psi_f(\vec r=0)\int M_0(\vec r)d^3 r=(2\pi)^{3/2}\psi_f(\vec r=0) M_0(\vec p=0)\,.
\label{eq:rf2}
\end{equation}
This relation takes place not only for bound states but also for the creation of   $h^+h^-$ pair in the continuous state if $kr_s\ll 1$.

The amplitude  $R_f$ is normalized by the relation
\begin{equation}
(2\pi)^32E_f\frac{d\sigma}{d\vec p_f}\Biggr|_{\vec p_f=0}=\frac{1}{4\sqrt{(p_+ p_-)^2-m_+^2m_-^2}}
\int \frac{\left| R_f\right|^2\Delta f}{2\mu(2\pi)^3}d\Phi_x\,,
\end{equation}
where $\Delta f=1$ for discrete   states  and $\Delta f=k^2\Delta k/2\pi^2$ in the continuous case.

Substituting the approximate relation (\ref{eq:rf2}) in this expression and making use
of the definition (\ref{eq:cs}) of the double differential cross section, one gets

\begin{equation}
2E_f\frac{d\sigma}{d\vec p_f}\Biggr|_{\vec p_f=0}= \frac{(2\pi)^3\,\left| \psi_f(\vec
r=0)\right|^2\Delta f}{\mu}E_+E_-\frac{d\sigma_0}{d\vec p_+d\vec p_-}\biggl|_{\vec
p_+=\vec p_-=0}\,,
\label{eq:2ef}
\end{equation}
where $\frac{d\sigma_0}{d\vec p_+d\vec p_-}\biggl|_{\vec p_+=\vec p_-=0}$  is the double differential  pair production cross section without the final state interaction.
As the combinations
$$ E_f\frac{d\sigma}{d\vec p_f},\quad
E_+ E_-\frac{d\sigma_0}{d\vec p_+d\vec p_-}$$
are relativistic invariant, it follows from (\ref{eq:2ef}) that in any reference frame
\begin{equation}
\label{eq:cs}
2E_f\frac{d\sigma}{d\vec p_f}=\frac{(2\pi)^3\,\left| \psi_f(\vec r=0)\right|^2\Delta f}{\mu} E_+ E_-
\frac{d\sigma_0}{d\vec p_+d\vec p_-} \Biggr|_{\vec p_+=\alpha_+\vec p_f;\:\vec p_-=\alpha_-\vec p_f}\,,
\end{equation}
where $\alpha_+=m_+/(m_++m_-)$, $\alpha_-=1-\alpha_+=m_-/(m_++m_-)$.

This  expression connects in the closed form the inclusive production cross sections of
bound and unbound states of the interacting meson pair with the double inclusive production cross section of noninteracting pair at zero relative momentum. For the case of bound states it was
obtained by L.~Nemenov~\cite{nemenov85}. This expression allows one to calculate, in common approach, the relative  probabilities  of meson pairs production  in bound and free states, which is a key point for the DMA lifetime measurement~\cite{adeva11}.

\section{Relative probability of the bound and unbound meson pairs production}

The dimesoatoms lifetime measurement  in experiments DIRAC at CERN based on evaluation of a DMA breakup probability in the target, which is a ratio between measured number the broken atoms in the target to the number of produced atoms. The latter is extracted from the number of observed pairs in the free state using the ratio between pairs production  in bound and free states \cite{afan99}. Let us reconsider this ratio using equation (\ref{eq:cs}). A first conclusion is that
the relative probability of pairs production in different states  is determined  solely by a
 two-meson wave function in the  appropriate state.

Accounting  that the two-meson  wave function with relative orbital momenta  $L$
behaves at small relative  distances $r$ as   $\psi (r)\sim r^L $ only pairs in the S-state should be taken into account.
The Coulomb wave functions of discrete $n$S and continuous  $k$S states at zero relative distance   reads~\cite{LL}
\begin{eqnarray}
\label{eq:psi0}
|\psi_{nS}(\vec r=0)|^2 &=& \frac{1}{\pi}\left(\frac{\mu\alpha}{n}\right)^{3}\,,\nonumber\\
|\psi_{kS}(\vec r=0)|^2 &= &C^2(k)=\frac{\pi\xi}{sh(\pi\xi)}e^{\pi\xi}=\frac{2\pi\xi}{1-e^{-2\pi\xi}}\,.
\end{eqnarray}

At large relative momentum $k\gg \mu\alpha$ $(\xi =\mu\alpha/k)$  the distribution
of interacting and noninteracting pairs coincides as $C(k)\to 1$.
From the other hand, at small momentum  $k\leq \mu\alpha$ $(\xi \geq 1)$  one can neglect the exponential term in (\ref{eq:psi0}) with the result:
\begin{equation}
 |C(k)|^2\simeq
2\pi\xi=\frac{2\pi\mu\alpha}{k}\,.
\end{equation}

Thus, for small relative momenta the Coulomb interaction in the final state modifies significantly
the distribution for the opposite charged hadrons
compared to noninteracting one changing it from relative momentum independence   to pole behavior:
\begin{equation}
\frac{d^3\sigma}{d^3\,k}\Biggr|_{\mid\vec k\mid\to 0}\to
\frac{const}{|\vec k|}\,.
\end{equation}

The production of unbound pairs with small relative momenta   $r\leq k_0= 2\mu\alpha$ is the main
background in extraction of DMA signal from experimental data. The Coulomb interaction in continuous spectra leads to the huge value of this background as compared with the production of  noninteracting meson pairs:
\begin{equation}
R_c=\frac{\int\limits_{0}^{k_0}C^2(k)k^2dk}{\int\limits_{0}^{k_0}k^2dk}\simeq
\frac{6\pi\mu\alpha}{k_0}\simeq 10\, .
\end{equation}

Making use the above equations, one can estimate  the relative probabilities of DMA
production in different $n$S states and the relative probability of unbound pairs production
in comparison with the dimesoatom production in the ground state:
\begin{eqnarray}
\label{eq:ns}
\frac{w_{nS}}{w_{1S}}&=&\frac{\left| \psi_{nS}(\vec r=0)\right|^2}
{\left| \psi_{1S}(\vec r=0)\right|^2}=
\frac{1/\pi\left(\mu\alpha/n\right)^3}
{1/\pi\left(\mu\alpha/1\right)^3}=\frac{1}{n^3}\,, \\
\label{eq:ratio}
\frac{dw_{kS}}{w_{1S}}&=&\frac{\left| \psi_{kS}(\vec
r=0)\right|^2} {\left| \psi_{1S}(\vec r=0)\right|^2}\cdot
\frac{k^2dk}{2\pi^2}=
\frac{kdk}{(\mu\alpha)^2\left(1-\exp(-2\pi\mu\alpha/k)\right)}\,.
\end{eqnarray}

The latter expression coincides with the well-known Gamov--Sommerfeld--Sa\-kh\-arov factor
\cite{GAMO28, SOMM31, SAKH91} derived for production of different pairs of oppositely charged particles. For the case of the charged meson pairs production such formulas was obtained in \cite{nemenov85}.

\section{Accuracy of the ``zero radius'' approximation}

Thus far we exploit  the fact that the DMA  wave function  is a slow varying function in
comparison with the pair production amplitude managed by a short range  strong interaction.
To estimate the accuracy of this approximation,  let us consider  the amplitude of  DMA
production  in  the ground  state
\begin{equation}
 \label{eq:rs}
 R_{1S}=\int M_0(\vec r)\psi_{1S}(\vec r)d^{\,3} r=4\pi\int M_0(\vec r)\psi_{1S}(\vec r)r^2dr\, .
\end{equation}
Making use   the Coulomb wave function for the ground state
\begin{equation}
\psi_{1S}(\vec r)=\frac{1}{\sqrt{\pi}}\left(\mu\alpha\right)^{3/2}
\exp\left(-\mu\alpha r\right)
\end{equation}
and choosing  the  Yukava type  representation~\cite{dirac95}  for the  amplitude  of the free pairs creation\footnote{In the momentum space this
parametrization  corresponds to pole - like dependence: $M_0(p)=M_0(0)\frac{\kappa^2}{\kappa^2+p^2}$, where $\kappa$  is a free parameter.}
\begin{equation}
M_0(\vec r)=\sqrt{2\pi}M_0(\vec p=0)\kappa^2e^{-\kappa  r}/r\,,
\end{equation}
the amplitude (\ref{eq:rs}) can be calculated with the result:
\begin{eqnarray}
\label{eq:rs1}
R_{1S}(\kappa)=4\sqrt{2}\pi(\mu\alpha)^{3/2}M_0(\vec p=0)\frac{\kappa^2}{(\kappa+\mu\alpha)^2}\, .
\end{eqnarray}

The  ``zero radius''  approximation corresponds to the limit $\kappa\to\infty$.
Choosing the  parameter $\kappa$   from  the interval~\cite{dirac95} 80 MeV $\leq\kappa\leq$ 140 MeV, one obtains the estimation  for  the ratio of the DMA production probability
 calculated with exact expression (\ref{eq:rs1}) to the  approximate one calculated using ``zero radius'' approximation:
\begin{equation}
\frac{R^2_{1S}(\kappa)}{R^2_{1S}(\infty)}=
\left(\frac{\kappa}{\kappa+\mu\alpha}\right)^4=0.950\div 0.975.
\end{equation}
Thus, the account of corrections on finite radius of strong interaction can reduce the DMA
production cross section on $2.5\div 5.0\%$.

The above consideration is the way to estimate the influence of the strong interaction on the accuracy of the cross sections in the form (\ref{eq:cs}). However, for the ratio (\ref{eq:ratio}) used for the experimental data analysis, the relative effect of this influence is of order of  $10^{-3}$ only \cite{afan99}.

Let us generalize the above consideration to the case of any $n$S  states.
The DMA wave function  for any  $n$S state reads~\cite{LL}
\begin{equation}
\psi_{nS}(\vec r)=\frac{1}{\sqrt{\pi}}\left(\frac{\mu\alpha}{n}\right)^{3/2}
\exp\left(-\frac{\mu\alpha}{n} r\right) F\left(1-n,2, \frac{2\mu\alpha}{n}r\right)\,,
\end{equation}
where  $ F\left(1-n,2, 2\mu\alpha r/n\right)$ is the confluent  hypergeometric function.

To determine  the probability of DMA production to any $n$S state one should calculate the
relevant amplitude
\begin{equation}
\label{eq:rns}
R_{nS}(\kappa)=\int M_0(\vec r)\psi_{nS}(\vec r)d^{\,3} r\,.
\end{equation}
To calculate this  amplitude,   one should   compute  the integral
\begin{eqnarray}
\label{eq:int}
 I&=&\int\limits_{0}^{\infty}e^{-\lambda r}r F(1-n,2,\omega r)dr\,,\\
 \lambda&=&\kappa+\mu\alpha/n,\quad \omega=2\mu\alpha/n \,.\nonumber
\end{eqnarray}
The general form of this integral is cited in the Appendix with  the result
\begin{equation}
I=\frac{1}{\lambda^2}\left (1-\frac{\omega}{\lambda}\right )^{n-1}\,.
\end{equation}

The amplitude (\ref{eq:rns}) for any $n$S state production can be presented in the form
\begin{eqnarray}
R_{nS}(\kappa)=4\sqrt{2}\pi(\frac{\mu\alpha}{n})^{3/2}M_0(\vec p=0)\frac{\kappa^2}{(\kappa+\mu\alpha/n)^2}\left(\frac{\kappa n-\mu\alpha}{\kappa n+\mu\alpha}\right )^{n-1}\,.
\end{eqnarray}
This  expression allows one to estimate the corrections to the ``zero radius'' approximation for
any $n$S state  production  in proton-proton collisions.

\section{Amplitude of DMA transition from  1S to 2P  state}
Let us consider the inelastic transition from the DMA ground state
1S to the first bound state with nonzero orbital, i.e. 2P state. The selection
rules allow only transitions to the states with $|m|=1$ (transition to
the state with $m=0$ is forbidden). The amplitude for such a transition
reads
\begin{eqnarray}
A_{fi}(\vec q)&=&\int d^2s f(\vec q,\vec s)h_{fi}(\vec s)\,, \nonumber\\
f(\vec q,\vec s)&=&\frac{i}{2\pi}\int d^2b\left[1-e^{i\Delta\chi(\vec b,\vec s)}\right]
 e^{i\vec q\vec b}\,,\nonumber\\
h_{fi}(\vec s)&=&\int\limits_{-\infty}^{\infty}dz\psi_f^{\ast}(\vec r)\psi_i(\vec r)\,, \quad \vec r=(\vec s,z)
\label{eq:hfi}
\end{eqnarray}
with  the wave functions $\psi_{i(f)}(\vec r)$ of the initial  $(|i\rangle
=|1S\rangle$ and final  $|f\rangle=|2P^{(\pm 1)}\rangle)$ states:
\begin{eqnarray}
\psi_i(\vec r)&=&\frac{(\mu\alpha)^{3/2}}{\sqrt{\pi}}
e^{-\mu\alpha r}\,,\nonumber\\
 \psi_f(\vec r)&=&\frac{(\mu\alpha)^{5/2}\vec\epsilon_+
 \vec r}{4\sqrt{2\pi}}e^{-\mu\alpha r/2}\,,
 \quad \vec \epsilon_\pm=\vec e_x\pm i\vec e_y\,.
\end{eqnarray}

Substituting these expressions in (\ref{eq:hfi}), we obtain
\begin{equation}
 h_{fi}(\vec s)=\frac{(\mu\alpha)^{4}}
{2\sqrt{2}\pi}(\vec \epsilon_+\vec s)sK_1(\tilde \mu s)\,,\quad
\tilde \mu=\frac{3\mu\alpha}{2}\,.
\end{equation}

On the other hand, due to wave functions orthogonality we have
\begin{equation}
\int h_{fi}(\vec s)d^2s=0\,.
\end{equation}

At small transfer momenta $q^2\ll \tilde \mu^2$ only single photon exchange is essential.
In this case the integration
in (\ref{eq:hfi}) can be done with the result
\begin{eqnarray}
 A_{fi}(\vec q)&=&A_{fi}^{1B}(\vec q)\left\{1+O\left[\frac{q^2}{\tilde\mu^2}
\ln\left(\frac{\tilde\mu^2}{q^2}\right)(Z\alpha)^2\right] \right\}\,,\nonumber\\
 A_{fi}^{1B}(\vec q)&=&i\alpha Z\sqrt{2}(\mu\alpha)^4
\frac{(\vec \epsilon_{\mp}\vec q)}{q^2} \cdot\tilde
\mu\left(\tilde\mu^{-1}\frac{\partial}{\partial\tilde\mu}\right)^2
\frac{1}{\tilde\mu^{2}+q^2/4}\,.
\end{eqnarray}
In this expression the factor $(\vec \epsilon_{\mp}\vec q)/q^2$ appears as we neglect the screening effect considering
the transition $(1S\to 2P^{\pm 1})$ at transfer momentum $q$ much larger than the inverse Born
radius of the target atoms $ \lambda_B=1/R_{sc}\approx m_e \alpha Z^{1/3}$.

 The Born amplitude generalized to take into account the screening effect takes the form
\begin{eqnarray}
 A_{fi}^{1B}(\vec q)&=&i\alpha Z\sqrt{2}(\mu\alpha)^4\frac{(\vec\epsilon_{\mp}\vec q)}{q^2+\lambda_B^2}
\mu\left(\tilde\mu^{-1}\frac{\partial}{\partial\tilde\mu}\right)^2
\frac{1}{\tilde\mu^{2}+q^2/4}\nonumber\\
&=&i\frac{3\sqrt{2}}{4} (4\mu\alpha)^5 Z\frac{(\vec\epsilon_{\mp}\vec q)}{(q^2+\lambda_B^2)((3\mu\alpha)^2+q^2)^3}\,.
\label{eq:aif1b}
\end{eqnarray}
The differential cross section  of the dimesoatom transition from 1S to 2P state is connected with the relevant amplitude (\ref{eq:hfi}) by a standard  relation
\begin{eqnarray}
\frac{d\sigma}{d^2q}=| A_{fi}(\vec q)|^2\,.
\end{eqnarray}
Substituting the  inverse Born radius of the target atoms in (\ref{eq:aif1b}) and summing the square of Born amplitude (\ref{eq:aif1b}) by final state polarization we obtain
the dimesoatoms transition cross section as a function of  transfer momenta  in Born approximation (single photon exchange)
\begin{eqnarray}
 d\sigma=\frac{9}{8}Z^2(4\mu\alpha)^{10}\frac{q^2d^2q}{\left(q^2+(m_e\alpha Z^{1/3})^2\right)^2\left(q^2+(3\mu\alpha)^2\right)^6}\,.
\end{eqnarray}
As mentioned above, this expression is valid in the wide interval of transfer momenta  giving the main contribution in the yield of 2P states of dimesoatoms, as a result of 1S state dimesoatoms  interaction with atoms of the target.

\section{Summary}
In this paper we consider the production of oppositely
charged meson pairs in collision of relativistic particles and obtain
general expressions for amplitudes and cross sections for the
creation of any bound states (dimesoatoms) and continuous states
accounting for the electromagnetic interaction in the final state. We
derive the general expression for such type production using a
coordinate representation for relevant wave functions and amplitudes.
We obtain the expressions which allow one to estimate the relative
probability of bound and continuous states production for any mesons
pair. The amplitude for any bound state production beyond the widely
used in the literature ``zero radius'' approximation has been obtained.
Finally, we obtain the analytical expression for the cross section of
dimesoatom transition from 1S to 2P state that is essential for the
pionium Lamb shift measurements.

\section*{Appendix}

To calculate the integral
$ I=\int\nolimits_{0}^{\infty}e^{-\lambda r}r^{b-1}F(a,b;\omega r)dr\,,$
it is convenient to use the representation of the confluent hypergeometric function in the form of contour integral
\begin{equation}
 F(a,b;z)=-\frac{1}{2\pi i}\frac{\Gamma(1-a)\Gamma(b)}{\Gamma(b-a)}\oint\limits_{C}e^{tz}(-t)^{a-1}(1-t)^{b-a-1}dt\,.
\end{equation}
The contour C begin  at $t=1$, bypass the point $t=1$ counterclockwise and return at point $t=1$.

This representation allows one to carry on the integration  with the result
\begin{eqnarray}
I=-\frac{1}{2\pi i}\frac{\Gamma(1-a)\Gamma(b)^2}{\Gamma(b-a)\lambda^{b}}\oint\limits_{C}
\frac{(-t)^{a-1}(1-t)^{b-a-1}}{(1-\omega
t/\lambda)^{b}}dt\nonumber\\
=\frac{\Gamma(b)^2}{\lambda^{b}}~_2F_1\left(a,b;b;\frac{\omega}{\lambda}\right)\,,
\end{eqnarray}
where the hypergeometric Gauss function $_2F_1(a,b;c;z)$ is determined by the series
\begin{eqnarray}
_2F_1(a,b;c;z)&\equiv & F(a,b;c;z)=\sum_{j=0}^{\infty}\frac{(a)_j\cdot (b)_j}{(c)_j\cdot j!}z^j\,,\nonumber\\
(a)_j&=&\frac{\Gamma(a+j)}{\Gamma(a)}=a(a+1)\ldots(a+j-1)\,,\nonumber\\
(b)_j&=&\frac{\Gamma(a+j)}{\Gamma(b)}=b(b+1)\ldots(b+j-1)\,,\nonumber\\
(c)_k&=&\frac{\Gamma(a+j)}{\Gamma(c)}=c(c+1)\ldots(c+j-1)\,,\nonumber\\
(a)_0&=&(b)_0=(c)_0=1\,,
\end{eqnarray}
or with the representation through the contour integral
\begin{equation}\label{43}
F(a,b;c;z)=-\frac{1}{2\pi i}
\frac{\Gamma(1-a)\Gamma(c)}{\Gamma(c-a)} \oint\limits_{C}(-t)^{a-1}(1-t)^{c-a-1}(1-tz)^{-b}dt\,.
\end{equation}
The contour $C$ in (\ref{43}) is the same as in the definition of the confluent hypergeometric function. Substitution $t\to t/(1-z+zt)$ in the integral (\ref{43}) leads to the relation between the hypergeometric functions at different  values of variables $z$ and  $z/(z-1)$:
$$F(a,b;c;z)=(1-z)^{-a}F\left(a,c-b;c;\frac{z}{z-1}\right)\,.$$

Applying this relation to the hypergeometric function,   one obtains
$$F\left(a,c;c;\frac{\omega}{\lambda}\right)=
\left(1-\frac{\omega}{\lambda}\right)^{-a}F\left(a,0;c;
\frac{\omega}{\omega-\lambda}\right)\,,$$
where $F(a,0;c;z)=1$ as $(b)_j=0$ at $j\ne 0$ and $b=0$ in definition of hypergeometric function
by series.

\end{document}